\documentclass[traditabstract, letter]{aa}  
\usepackage{amsmath}
\usepackage{amssymb}
\usepackage{natbib}
\usepackage{graphicx}
\usepackage{txfonts}
\usepackage[english]{babel}
\usepackage{units}
\begin{document}
\title{The lunar phases of dust grains orbiting Fomalhaut}


\author{
	M. Min\inst{1}
		\and
	M.~Kama\inst{2}
		\and
	C.~Dominik\inst{2, 3}
		\and
	L.~B.~F.~M.~Waters\inst{2, 4}
}

\offprints{M. Min, \email{M.Min@uu.nl}}

\institute{
Astronomical Institute Utrecht, University of Utrecht, P.O. Box 80000, NL-3508 TA Utrecht, The Netherlands
	\and
Astronomical Institute Anton Pannekoek, University of Amsterdam,
Kruislaan 403, 1098 SJ  Amsterdam, The Netherlands
	\and
Afdeling Sterrenkunde, Radboud Universiteit Nijmegen,
Postbus 9010, 6500 GL Nijmegen, The Netherlands
	\and
Insituut voor Sterrenkunde, K.U.Leuven,
Celestijnenlaan 200 D, B-3001 Leuven, Belgium
}

   \date{Received August 5, 2009; accepted December 18, 2009}

 
  \abstract
{Optical images of the nearby star Fomalhaut show a ring of dust orbiting the central star. This dust is in many respects expected to be similar to the zodiacal dust in the solar system. The ring displays a clear brightness asymmetry, attributed to asymmetric scattering of the central starlight by the circumstellar dust grains. Recent measurements show that the bright side of the Fomalhaut ring is oriented away from us. This implies that the grains in this system scatter most of the light in the backward direction, in sharp contrast to the forward-scattering nature of the grains in the solar system. In this letter, we show that grains considerably larger than those dominating the solar system zodiacal dust cloud provide a natural explanation for the apparent backward scattering behavior. In fact, we see the phases of the dust grains in the same way as we can observe the phases of the Moon and other large solar system bodies. We outline how the theory of the scattering behavior of planetesimals can be used to explain the Fomalhaut dust properties. This indicates that the Fomalhaut dust ring is dominated by very large grains. The material orbiting Fomalhaut, which is at the transition between dust and planetesimals, can, with respect to their optical behavior, best be described as micro-asteroids.
}


   \maketitle
%

\section{Introduction}

Optical images of some nearby stars show the presence of orbiting solid material through the starlight it scatters \citep[e.g][]{2005Natur.435.1067K}.
This has been interpreted as evidence of an evolving planetary system. Collisions between large bodies such as asteroids are believed to be responsible for the production of these small solid particles. This is similar to our own solar system, in which interplanetary dust grains are found that are thought to originate from asteroids and comets. The size and chemical composition of these grains determine the way in which they reflect starlight. 
Laboratory measurements and light scattering theory show that small solid particles scatter most incident light in the forward direction, in agreement with astronomical observations.
However, recent observations of the dust orbiting the nearby star Fomalhaut have shown for the first time that the grains in this system scatter most of the light in the backward direction \citep{2009A&A...498L..41L}. This is in sharp contrast with the current understanding of light scattering and challenges the analogy between interplanetary dust and the dust observed in Fomalhaut. 
In this paper, we show that an excellent match to the observed backscattering behavior of the material orbiting Fomalhaut can be obtained by using the theory of regolith covered surfaces.

\section{Apparent backward scattering}

The circumstellar ring of Fomalhaut as observed by the Hubble space telescope clearly shows a dark side and a bright side \citep{2005Natur.435.1067K}. Regarding the orientation of the ring, the general forward scattering behavior of dust grains has led to the conclusion that the bright side of the ring should be the one tilted towards the observer.
Using the Very Large Telescope Interferometer (VLTI), the rotation axis of the central star and subsequently the inclination axis of the system was inferred \citep{2009A&A...498L..41L}. This inclination is opposite to previous assumption in that the direction of scattering is reversed from forward-dominated to backward-dominated. This contrasts with the current consensus in the field of light scattering.

A relatively simple explanation for this can be found by considering bodies much larger than interplanetary dust grains. As an example let us take the Moon, which for all practical purposes can be considered a gigantic dust grain. However, it is evident that the Moon is bright when viewed in a backward scattering situation (full Moon) and it is dark in the forward scattering case (new Moon). The question then arises: is the Moon truly backscattering? The answer is ``no''. The apparent surprise that the Moon is forward scattering is caused by the sharp and in most cases invisible diffraction peak it causes. For the Moon this diffraction spike, caused by interference effects due to the shadow of the Moon, is so narrowly forward peaked that it can only be seen in the forward-most fraction of a degree\footnote{For objects the size of the Moon, the diffracted energy at visible wavelengths is confined to within the forward-most $\mu$-arcsecond.}, i.e. when the Sun is right behind the Moon. For all practical purposes, the Moon appears to be backscattering, and diffraction can be omitted.

\begin{figure}[!t]
\centerline{\resizebox{\hsize}{!}{\includegraphics{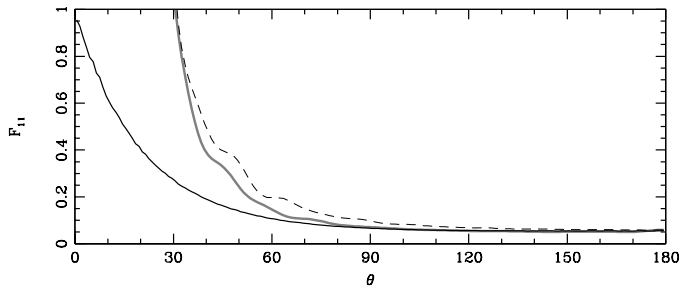}}}
\caption{Angular scattering functions for the regolith particles, showing the relative intensity scattered at a scattering angle $\theta$. The pure reflectance function (solid, black curve) is given, and diffraction effects are added for particles with a narrow size distribution around 3$\,\mu$m grains (dashed curve). The exact scattering function for this narrow size distribution of spheres as obtained from the Mie theory is also shown (solid, grey curve).}
\label{fig:singlescatt}
\end{figure}

For smaller grains, the diffraction spike often dominates the overall scattering behavior over a broader range of scattering angles. It appears that the disk around Fomalhaut contains dust grains that behave backward-reflecting like asteroidal bodies. This must be because the grains are very large, making the diffraction spike so narrowly forward-peaked that it is outside the observable range of angles. For the Fomalhaut system, which has an inclination of $25^\circ$, the range of observable scattering angles is $25^\circ<\theta<155^\circ$. \citet{2005Natur.435.1067K} already noted that the apparent single scattering albedo of the dust grains is in the range of $5-10$\%, which is low for any kind of dust grains. Such values are more typical for the geometric albedo of asteroids. The geometric albedo of asteroids does not include the contribution from diffraction. For large grains, it is easily shown that when diffraction is included, the single scattering albedo is always larger than 50\% \citep[][Chapter 12]{vandeHulst}. Thus, the true single scattering albedo of the grains, averaged over all angles and including diffraction, must be much higher than the observed value, indicating that most of the scattered light is not detected.

\section{Modeling the scattering function of asteroidal bodies}

\begin{figure*}[!t]
\centerline{\resizebox{0.7\hsize}{!}{\includegraphics{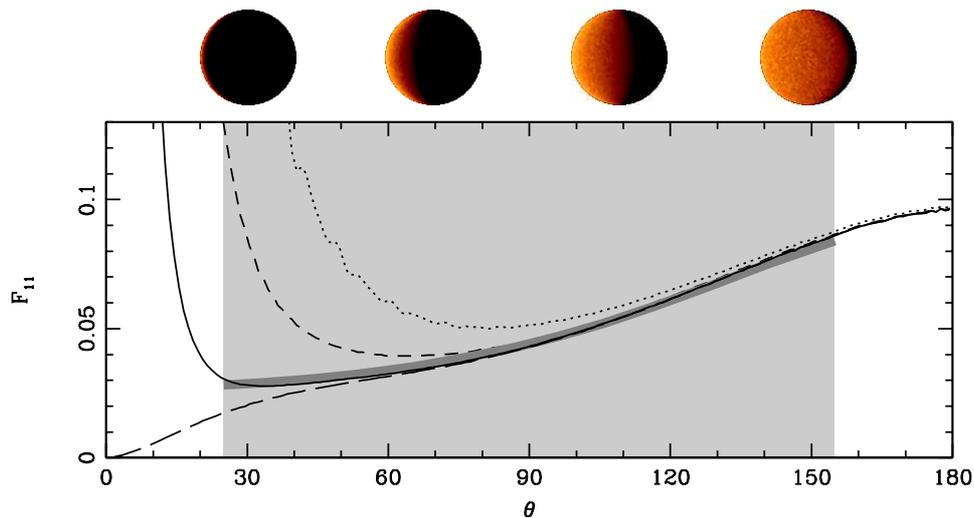}}}
\caption{Angular scattering functions for large dust grains, showing the relative intensity scattered at a scattering angle $\theta$. The pure reflectance function (long-dashed curve) is given, and diffraction effects are added for 10, 30 and 100$\,\mu$m grains (dotted, dashed and solid curves). The empirical scattering function of the Fomalhaut disk grains from \citet{2005Natur.435.1067K} is also shown (thick grey curve, albedo chosen to match the computations). The grey area indicates the observable range of scattering angles for the Fomalhaut system.}
\label{fig:F11}
\end{figure*}

Below we treat the dust grains in the Fomalhaut system as micro-asteroids, meaning that we use the theory of reflectance by regolith-covered asteroidal surfaces to compute their scattering behavior.
For the reflectance of asteroidal surfaces, we use an analytical model for the bidirectional reflectance \citep{1981JGR....86.3039H}. For simplicity we use the theory omitting effects of macroscopic roughness as discussed in \citet{1984Icar...59...41H}. This basically means that we set the macroscopic roughness parameter as discussed in that paper to zero. In addition to this, we ignore the opposition effect. This mainly plays a role near backward-scattering and thus largely falls outside the range of scattering angles that we have access to in the Fomalhaut system. To integrate over the surface of the body, we use Monte Carlo ray tracing.

\subsection{The regolith particles}

As input in these computations, we need to know the albedo and angular scattering function for the regolith particles that cover the surface of the grain. Note that since the regolith particles are closely packed, we need to know these parameters omitting the contribution from diffraction \citep{1981JGR....86.3039H}. The contribution of diffraction can be separated when we compute the angular scattering function and single scattering albedo using geometrical optics. More exact computational methods like the Mie theory do not allow the separation of the contribution from diffraction. We thus assume the regolith particles are large enough so that geometrical optics can be applied, and in addition we only consider surface reflections, i.e. transmitted rays are ignored. We realize that using geometrical optics is a rough approximation for the grain sizes we are considering. Nevertheless, since we are interested only in the overall shape of the angular scattering function mostly at intermediate scattering angles, the approximation is not so bad for grain sizes a few times the wavelength of radiation, especially for particles with a considerable imaginary part of the refractive index \citep{1997ApOpt..36.4305W}. In this approximation the scattering properties of the particles are independent of their size and shape as long as they have a convex surface. To compute the reflections we need to specify the refractive index of the regolith particles which we take to be $m=1.6+0.1i$, thought to be typical for cosmic materials. However, we find that the final result is hardly influenced by modest variations in the refractive index (with the real part $\sim1.5-1.8$ and the imaginary part $\lesssim0.3$). The angular scattering function for the regolith particles which results from this is dominantly forward-scattering, while their single scattering albedo is $11$\% (omitting diffraction). In Fig.~\ref{fig:singlescatt} we show three different curves: the angular scattering function omitting diffraction (solid, black curve), the angular scattering function with added contribution from diffraction by a narrow size distribution of spheres with diameters around $3\,\mu$m (dashed curve), and the full angular scattering function computed using the Mie theory for this narrow size distribution (solid, grey curve). All curves are plotted as a function of the scattering angle, i.e. $0^\circ$ is forward-scattering while $180^\circ$ is backward-scattering. Note that the curve obtained using reflection plus diffraction is quite similar to that obtained using the full Mie theory. It can be argued that 3$\,\mu$m is rather large for the regolith particles covering the grains we discuss below. However, it does allow the grains to be covered by or composed of these regolith particles. The parameters derived above are used as input for the optical properties of the regolith particles in the Hapke theory.

\subsection{The grains as a whole}

The resulting single scattering properties of the regolith particles obtained in the previous paragraph are inserted into the Hapke theory. Note that we use the theory in its most simple form, i.e. we ignore effects of macroscopic roughness and the opposition effect. In that case, the only remaining parameters are the single scattering albedo and the angular scattering function of the regolith particles obtained above. As we will show below, we can find an almost perfect fit without considering additional parameters on macroscopic roughness and opposition effect. This means that using only the measurements available so far, we cannot constrain them and decided to use the most simple form that can still reproduce the observations. Much more accurate observations of the angular scattering function of the Fomalhaut grains are needed to compare all Hapke parameters to those derived for, for example, the Moon or solar system asteroids.

In Fig.~\ref{fig:F11} we show the resulting angular scattering function of the model Fomalhaut dust grains. The curves show the relative intensity scattered at an angle $\theta$. The grey area indicates the observable range of scattering angles for the Fomalhaut system. Note that the grains we simulate using the Hapke theory appear to be predominantly backward-scattering (we see the crescents of the grains), while the small regolith particles covering the surface of these larger grains are predominantly forward-scattering, as discussed above. The angular scattering function resulting from only using the Hapke reflectance theory, i.e. without including diffraction, is shown by the long dashed line. To simulate the finite size of the dust grains, we added the contribution from diffraction using a diameter of the grains of $10$, $30$ and $100\,\mu$m in dotted, dashed and solid lines, respectively. For the diffraction, we took simple Fraunhofer diffraction by a spherical aperture. To get rid of the resonance structures associated with Fraunhofer diffraction, averaging was performed over a narrow size distribution (a flat distribution from 0.8 to 1.2 times the average size). Grains of $100\,\mu$m diameter clearly fit the best to the empirical angular scattering function (shown by the thick gray line) in agreement with \citet{2000MNRAS.314..702D}.
For the observed scattering function of the grains orbiting Fomalhaut we used the Henyey-Greenstein parameterization obtained by \citet{2005Natur.435.1067K}, where we simply switched the forward and backward scattering directions, i.e. switching the asymmetry parameter $g$ from $+0.2$ to $-0.2$. The angular scattering function obtained in this way was multiplicatively scaled to match the computations. This scaling factor gives the single scattering albedo of the grains. We found that when we integrated the angular scattering function resulting from the Hapke theory over the angles accessible in the Fomalhaut system (the area indicated in gray), the albedo of the grains averaged over the observable range of angles is $5$\%. This is consistent with the low albedo found by \citet{2005Natur.435.1067K}.

If we consider the assumptions on the regolith particles and the Hapke theory it is clear that the model we present is quite rough. However, a firm outcome of the model is that for the range of scattering angles we observe the scattering is dominated by reflection rather than by diffraction. This implies, as shown above, that the grains are larger than 100$\,\mu$m in diameter. This outcome does not depend on the assumptions in the model and is therefore a solid conclusion.

\section{Conclusion}

We conclude that the scattering surface in the disk around Formalhaut is dominated by grains of at least $100\,\mu$m in size. 
Our findings agree with the fact that the infrared spectrum is featureless \citep{2004ApJS..154..458S}, indicating that grains smaller than a few micron are heavily depleted in the system. Also, \citet{2000MNRAS.314..702D} find that large grains dominate the thermal emission.
It is remarkable that such large grains are observed directly in optical light in an astronomical object. The reason for this is that while the dust mass in many systems is dominated by large grains \citep[see e.g.][]{2005ApJ...626L.109W, 2003A&A...403..323T}, they are normally over-shone by a component of small grains. This component, though less massive, dominates the optical cross section, and therefore appearance, of the disk. The fact that very large grains are so visible on a global scale in the Formalhaut disk means that small grains have been cleaned out from this disk extremely efficiently, and that all there is left are these very large grains, directly probed by reflected stellar light.

We conclude that we are observing the transition part of parameter space, going from dust scattering to planetesimal reflection. In fact, one might conclude that, almost 400 years after Galileo observed the crescent of Venus, we are seeing the crescents of the large dust grains in the disk around Fomalhaut.

\begin{acknowledgements}
We are indebted to the referee, Ludmilla Kolokolova, for constructive comments leading to a significant improvement of the paper.
\end{acknowledgements}


\begin{thebibliography}{10}
\expandafter\ifx\csname natexlab\endcsname\relax\def\natexlab#1{#1}\fi

\bibitem[{{Dent} {et~al.}(2000){Dent}, {Walker}, {Holland}, \&
  {Greaves}}]{2000MNRAS.314..702D}
{Dent}, W.~R.~F., {Walker}, H.~J., {Holland}, W.~S., \& {Greaves}, J.~S. 2000,
  \mnras, 314, 702

\bibitem[{{Hapke}(1981)}]{1981JGR....86.3039H}
{Hapke}, B. 1981, \jgr, 86, 3039

\bibitem[{{Hapke}(1984)}]{1984Icar...59...41H}
{Hapke}, B. 1984, Icarus, 59, 41

\bibitem[{{Kalas} {et~al.}(2005){Kalas}, {Graham}, \&
  {Clampin}}]{2005Natur.435.1067K}
{Kalas}, P., {Graham}, J.~R., \& {Clampin}, M. 2005, \nat, 435, 1067

\bibitem[{{Le Bouquin} {et~al.}(2009){Le Bouquin}, {Absil}, {Benisty}, {Massi},
  {M{\'e}rand}, \& {Stefl}}]{2009A&A...498L..41L}
{Le Bouquin}, J.-B., {Absil}, O., {Benisty}, M., {et~al.} 2009, \aap, 498, L41

\bibitem[{{Stapelfeldt} {et~al.}(2004){Stapelfeldt}, {Holmes}, {Chen}, {Rieke},
  {Su}, {Hines}, {Werner}, {Beichman}, {Jura}, {Padgett}, {Stansberry},
  {Bendo}, {Cadien}, {Marengo}, {Thompson}, {Velusamy}, {Backus}, {Blaylock},
  {Egami}, {Engelbracht}, {Frayer}, {Gordon}, {Keene}, {Latter}, {Megeath},
  {Misselt}, {Morrison}, {Muzerolle}, {Noriega-Crespo}, {Van Cleve}, \&
  {Young}}]{2004ApJS..154..458S}
{Stapelfeldt}, K.~R., {Holmes}, E.~K., {Chen}, C., {et~al.} 2004, \apjs, 154,
  458

\bibitem[{{Testi} {et~al.}(2003){Testi}, {Natta}, {Shepherd}, \&
  {Wilner}}]{2003A&A...403..323T}
{Testi}, L., {Natta}, A., {Shepherd}, D.~S., \& {Wilner}, D.~J. 2003, \aap,
  403, 323

\bibitem[{van~de Hulst(1957)}]{vandeHulst}
van~de Hulst, H.~C. 1957, Light Scattering by Small Particles (New York: Wiley)

\bibitem[{{Wielaard} {et~al.}(1997){Wielaard}, {Mishchenko}, {Macke}, \&
  {Carlson}}]{1997ApOpt..36.4305W}
{Wielaard}, D.~J., {Mishchenko}, M.~I., {Macke}, A., \& {Carlson}, B.~E. 1997,
  \ao, 36, 4305

\bibitem[{{Wilner} {et~al.}(2005){Wilner}, {D'Alessio}, {Calvet}, {Claussen},
  \& {Hartmann}}]{2005ApJ...626L.109W}
{Wilner}, D.~J., {D'Alessio}, P., {Calvet}, N., {Claussen}, M.~J., \&
  {Hartmann}, L. 2005, \apjl, 626, L109

\end{thebibliography}
\end{document}